\documentclass[preprint,aps,showpacs,,nofootinbib,superscriptaddress,preprintnumbers,epsf,psf]{revtex4}
\usepackage[english]{babel}
\usepackage{pstricks}
\usepackage[dvips]{graphicx}
\usepackage{epsf}
\usepackage{amsmath}
\usepackage{amsfonts}
\usepackage{amssymb}
\usepackage{epsf}
\usepackage[dvips]{graphicx}
\usepackage{dcolumn}
\usepackage{bm}
\everymath{\displaystyle}
\begin{document}

\title{Chiral vortaic effect and neutron asymmetries at NICA}

\author{\firstname{Oleg V.}~\surname{Rogachevsky}}
\email{rogachevsky@jinr.ru} \affiliation{JINR, 141980 Dubna
(Moscow region), Russia} \affiliation{PNPI RAS, 188300 Gatchina
(Leningrad district), Russia}
\author{\firstname{Alexander S.}~\surname{Sorin}}
\email{sorin@theor.jinr.ru}
\author{\firstname{Oleg V.}~\surname{Teryaev}}
\email{teryaev@theor.jinr.ru} \affiliation{JINR, 141980 Dubna
(Moscow region), Russia} \affiliation{Dubna International
University, 141980 Dubna (Moscow region), Russia}

\date{\today}

\begin {abstract}
We study the possibility of testing experimentally signatures of
P-odd effects related with the vorticity of the medium. The Chiral
Vortaic Effect is generalized to the case of conserved charges
different from the electric one. In the case of baryonic charge
and chemical potential such effect should manifest itself in
neutron asymmetries at the NICA accelerator complex measured by
the MPD detector. The required accuracy
may be achieved in a few months of accelerator running. We also
discuss polarization of the hyperons and P-odd correlations of
particle momenta (handedness) as probes of vorticity.

\begin{center}

 {\it \large We dedicate this paper to the memory of Academician

Alexei Norairovich Sissakian}

\end{center}

\end{abstract}


\maketitle

\section{Introduction}

The local violation \cite{Fukushima:2008xe} of discrete symmetries
in strongly interacting QCD matter is now under intensive
theoretical and experimental investigations. The renowned Chiral
Magnetic Effect (CME) uses the (C)P-violating (electro)magnetic
field emerging in heavy ion collisions in order to probe the
(C)P-odd effects in QCD matter.

There is an interesting counterpart of this effect, Chiral Vortaic
Effect (CVE)\cite{Kharzeev:2007tn,Kharzeev:RoundTable} due to
coupling to P-odd medium vorticity. In its original form
\cite{Kharzeev:2007tn} this effect leads to the appearance of the
same electromagnetic current as CME. Here we suggest a
straightforward generalization of CVE resulting in generation of
all conserved-charge currents. In particular, we address the case
of the {\it baryonic} charge and the corresponding asymmetries of
baryons, especially neutrons (not affected by CME), which can be
measured by the MultiPurpose Detector (MPD) \cite{CDR} at the
Nuclotron-based Ion Collider fAcility (NICA) \cite{nica} at the
Joint Institute for Nuclear Research (JINR).

\section{Chiral Magnetic and Vortaic effects}

The basic point in the emergence of CME is the coupling of the
topological QCD field $\theta$ \footnote{Its effect is potentially
much larger than the effects of CP-violation responsible e.g. for
neutron EDM.} to the electromagnetic field
$A_{\alpha}$ controlled by the triangle axial-anomaly diagram.
Similar interaction of $\theta$ with the velocity field
$V_{\alpha}$ exists in relativistic hydrodynamics
due to the new
coupling
\begin{eqnarray}
e_j A_{\alpha} J^{\alpha} \Rightarrow \mu_j V_{\alpha} J^{\alpha}
\end{eqnarray}
involving the chemical potentials $\mu_j$ (for various flavours
$j$) and the current $J^{\alpha}$. It provides also the
complementary description \cite{zakharov} of the recently found
contribution of fluid vorticity to the anomalous non-conserved
current \cite{Son:2009tf}. Note that the similarity between the
effects of the magnetic field and the rotation mentioned in
\cite{Kharzeev:2007tn} is very natural as the rotation is related
by the Equivalence Principle to the so called {\it
gravitomagnetic} field (see e.g. \cite{Obukhov:2009qs} and
references therein).

CVE leads to similar (to CME) contribution to the electromagnetic
current:
\begin{eqnarray}
\label{j} J^\gamma_e =  \frac{N_c}{4 \pi^2 N_f}
\varepsilon^{\gamma \beta \alpha \rho}
\partial_\alpha V_\rho
\partial_\beta (\theta \sum_{j} e_j \mu_j )~,
\end{eqnarray}
where $N_c$ and $N_f$ are the numbers of colours and flavours,
respectively. If variation of the chemical potential is neglected,
the charge induced by CVE for a given flavour can be obtained from
that due to CME by substitution of the magnetic field with the
curl of the velocity: $e_j \vec H \to \mu_j \vec\nabla \times \vec
V$.

In order to estimate the vorticity one may appeal
\cite{Kharzeev:2007tn} to the Larmor theorem relating the magnetic
field to the angular velocity of the rotating body, which in turn
is proportional to the vorticity. As a result, for $\mu \sim 500
MeV$ (in the NICA energy range) the order of magnitude of CVE
should be the same as that of CME.

On one hand, CVE provides another source for the observed
consequences of CME, relating with both light and strange
\cite{ot} quarks (regarded as the heavy ones
\cite{Polyakov:1998rb}). On the other hand (this is the basis of
our following discussion), CVE leads also to the separation of
charges different from the electric one. This becomes obvious if
the current is calculated from the triangle diagram, where quark
flavours $j$ carry various charges $g_{i(j)}$ (see Fig.1). The
calculation may  also be performed following
\cite{Kharzeev:2007tn} by variation of the effective Lagrangian
with respect to the external vector field. In that case this
vector field can be not only the electromagnetic potential
\cite{Kharzeev:2007tn} (entering the Lagrangian describing the
interaction with the real electromagnetic field) but also an
arbitrary (auxiliary) field coupled to any conserved charge.

If variation of the chemical potential in eq. (\ref{j}) is
neglected, the current of that charge $g_i$ selecting the specific
linear combination of quark triangle diagrams is related to
electromagnetic one as follows (see Fig.1):
\begin{eqnarray}
J_i^\nu =  \frac{\sum_{j}  g_{i(j)} \mu_j}{\sum_{j} e_j \mu_j}
~J_e^\nu~.
\end{eqnarray}
In another extreme case of dominance of chemical potential
gradients (assumed to be collinear) one gets the relation
\begin{eqnarray}
|J_i^0| =  \frac{|\vec\nabla\sum_{j}  g_{i(j)}
\mu_j|}{|\vec\nabla\sum_{j} e_j \mu_j|} |J_e^0|
\end{eqnarray}
which might be useful e.g. for the mixed phase \cite{nicaWP}
description.

\begin{figure}[t]
\centerline{\includegraphics[width=0.4\textwidth]{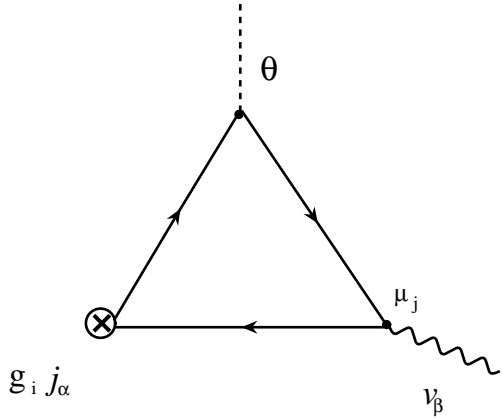}}
  \caption{The generation of the current of the conserved charge $g_i$ by the chemical potential $\mu_j$.}
\label{Fig-1}
\end{figure}

In particular, the large baryonic chemical potential (actually the
largest one which is achievable in accelerator experiments
\cite{Randrup:2006uz}), appearing in the collisions at
comparatively low energies at the FAIR and NICA (and possibly SPS
and RHIC at low energy scan mode) facilities, may result in the
separation of the baryonic charge. Of special interest are
manifestations of this separation in {\it neutron} asymmetries
with respect to the production plane, as soon as the neutrons,
from the theoretical side, are not affected by CME \footnote{Let
us stress that CME leads to the separation of all conserved
charges (including baryonic ones) of electrically charged
particles only.} and, from the experimental side, there is a
unique opportunity to study neutron production and asymmetries by
MPD at NICA. Besides that, the noticeable strange chemical
potential at the NICA energy range (see e.g. \cite{Toneev:2003sm}
and references therein) might result in the strangeness
separation.

\section{Experiments at NICA and neutron asymmetries}

The numerical smallness of such expected vortaic effect makes it
highly improbable to search it on an event-by-event basis. To
collect statistics from different events one needs to construct a
quadratic variable which does not depend on the varying sign of topological
field fluctuations.

This problem was solved in the experimental studies of CME
\cite{Voloshin:2004vk,Selyuzhenkov:2006fc,:2009txa,Voloshin:2010ju}
by consideration of the angular asymmetries of {\it pairs} of
particles with the same and opposite charges with respect to the
reaction plane. Moreover, one can use three-particle correlations
as well in order to avoid the necessity of fixing the reaction
plane.

We suggest to use the similar correlations for baryonic charge.
However, this method is not directly applicable in the case of
baryon charge separation because of the very small number of
produced antibaryons, in particular, antineutrons\footnote{Note
that the opposite sign charge correlations for CME are also very
small \cite{Voloshin:2010ju}.}. Nevertheless, the two-particle
correlation for neutrons still might be used as one of the probes
of CVE. In the case of three-particle correlations the third
particle should not necessarily be the neutron and could also be a
charged particle.

Note that the comparison of above-mentioned correlations for
various particles could be very useful. Namely, the direct effect of
CVE is negligible
for pions, due to the rather small chemical potential, so that
only CME contributes. On the other hand, for neutrons the
correlations are entirely due to CVE, while for protons one should
have both such effects. In case the correlations emerged due to
other reasons than CVE and CME to quadratic order, then their
simultaneous observation would be an important test of their
actual existence.


For the studies of CVE we suggest the collider NICA\footnote{The
value of CME at NICA is under intensive discussion \cite{nicaWP}.}
which is expected to operate with average luminosity $L \sim
10^{27} cm^{-2} s^{-1}$ for $Au + Au$ collisions in the energy
range $\sqrt{s_{N N}} = 4 \div 11$ {GeV/n} (for $Au^{79+}$ ). In
one of the collision points of NICA rings the Multi Purpose
Detector~(MPD)~\cite{CDR} will be located. MPD is proposed for a
study of dense baryonic matter in collisions of heavy ions over
the wide atomic mass range $A = 1 \div 197$. Inclusion of neutron
detectors is also considered in the conceptual design of the MPD.
The multiplicity of the neutrons in these collisions, predicted by
the UrQMD model \cite{UrQMD}, will be about $200$ in a full solid
angle. The number of registered neutrons in each event depends on
the event centrality and varies on the range $10 \div 150$ with a
reasonable efficiency $\sim 60\% $ for neutron detection. With the
proposed interaction rate for the detector MPD  of about
6~kHz~\cite{CDR}, it will be possible in a few months of
accelerator running time to accumulate $\sim 10^{9}$ events with
different centralities and measure CVE with comparable accuracy to
CME or set an upper limit on the value of CVE. For the estimation
of CVE we could explore the same three-particle correlator of azimuthal angles
\begin{eqnarray}
\label{corr}
\left< cos( \phi_\alpha + \phi_\beta - 2\phi_c) \right>
\end{eqnarray}
which was used for the detection of CME \cite{Voloshin:2004vk,
Selyuzhenkov:2006fc,:2009txa,Voloshin:2010ju}.

\begin{figure}[th]
\centerline{\includegraphics[width=0.45\textwidth]{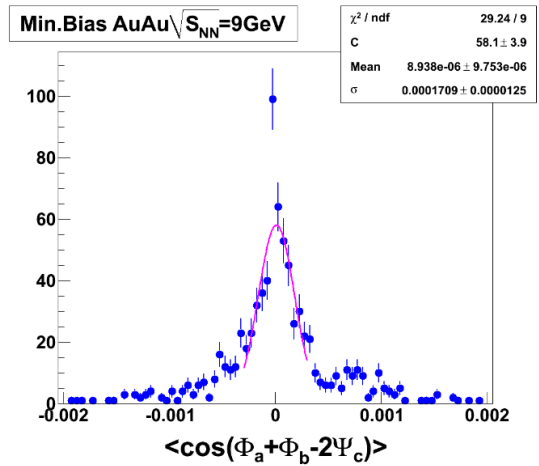}}
  \caption{Distribution of three neutrons correlator for Au+Au
  collisions at $\sqrt{s_{N N}} = 9$ {GeV/n} for UrQMD event
  generator. }
\label{cve-est}
\end{figure}

Fig.\ref{cve-est} shows the distribution of correlators
(\ref{corr}) for neutrons from UrQMD model events of minbias
$Au+Au$ collisions at $\sqrt{s_{N N}} = 9$ GeV. In each event the
correlator was obtained by taking two of the neutrons ($\alpha$
and $\beta$ in eq.  (\ref{corr})) from the mid-rapidity range
($|\eta| < 3$) and a third one (c in eq. (\ref{corr})) was taken
from the large rapidity range ($|\eta| > 3$). The correlators mean
value is equal to zero due to an absence of the neutron asymmetry
in the model simulation as it is shown on the Fig.\ref{cve-est}.

\begin{figure}[h]
\centerline{\includegraphics[width=0.5\textwidth]{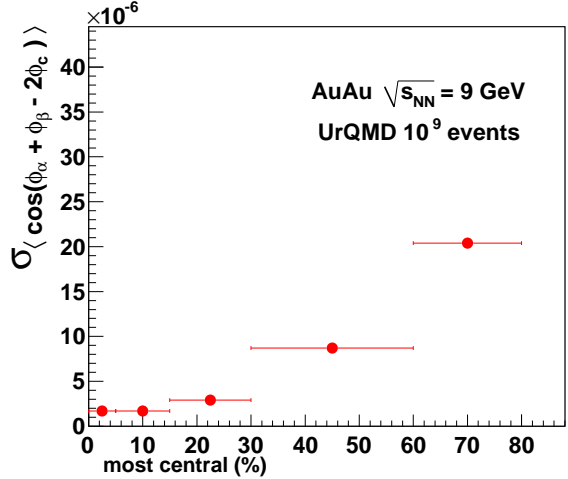}}
  \caption{Estimation of statistical errors. }
\label{Fig-2}
\end{figure}

We should mention that UrQMD model predicts the number of neutrons in
each event within the mid-rapidity range is much smaller than the
number of charged particles. Hence, in order to determine CVE with the
same value of precision as for CME case at RHIC
\cite{Voloshin:2010ju}, we need to have a much larger number of
events. For the rough estimation, while $\sim \mathrm 15\,M$ of events
were sufficient at RHIC for targeted precision in the CME case, at
NICA we need $\sim 1000\,M$ of events for the same precision in CVE
measurements, which could be accumulated in a few months of NICA/MPD
running time. The possible magnitude of the statistical errors for the
three-particle correlator with $10^{9}$ of collected events is shown
in Fig.\ref{Fig-2}.

At the moment only estimation of the statistical errors for the
correlator~(\ref{corr}) could be performed and more thorough
investigation of systematic and measurement errors (like detector
acceptance and inefficiencies, effects of particle clustering
etc.) should be carried out with obtained experimental data from
the real detector. It could be pointed out that the main
contribution to the background for correlator (\ref{corr}) comes
from particles flow and at the NICA energy range elliptic flow of
charged (and neutral) particles is less than at RHIC energies.
Therefore, one could expect that background of flow effect for
neutral particles used in CVE calculations should be also less
than that for charged particles in chiral magnetic effect at high
energy collisions. More detailed estimates taking into account
also neutron detector acceptances and efficiencies will be
discussed elsewhere.


\section{Conclusions and outlook}

We discussed the new tests of P-odd effects in heavy ions
collisions due to vorticity in the specific conditions of the MPD
detector at NICA. Special attention was paid to the generalization
of the Chiral Vortaic Effect to the case of separation of the
baryonic charge and its manifestation in neutron asymmetries.

We proposed to study the two- and three-particle correlations
similar to those used in studies of CME. We compared the required
accuracies and found that CVE could be studied with the data
collected in a few months of NICA running.

As an outlook, let us first mention that the non-perturbative (in
particular, lattice  QCD \cite{Buividovich:2009bh}) studies of
vorticity effects are very important. Let us also note that the
large chemical potential might result in meson decays forbidden in
the vacuum, like C-violating $\rho \to 2 \gamma$
\cite{Teryaev:1996dv,Radzhabov:2005jp} or recently considered
CP-violating $\eta \to 3 \pi$ \cite{Millo:2009ar}.

Vorticity is related to the global rotation of hadronic matter, an
interesting observable by itself.
Its calculations in the
framework of various models are very desirable, as well as studies
of its possible relations with other collective effects due to
non-centrality of heavy ion collisions, like directed ($v_1$) and
elliptic ($v_2$) flows.

Another interesting problem is the possible manifestation of
vorticity in the
polarization of $\Lambda$ particles was suggested some time ago in
\cite{Liang:2004ph} although the experimental tests at RHIC
\cite{Selyuzhenkov:2006fc} did not show any significant effect.
One may think that such a polarization can emerge due to the
anomalous coupling of vorticity to the (strange) quark axial
current via the respective chemical potential, being very small at
RHIC but substantial at FAIR and NICA energies. In that case the
$\Lambda$ polarization at NICA \cite{nicaWP} due to triangle
anomaly can be considered together with other probes of vorticity
\cite{Betz:2007kg} and recently suggested signals
\cite{KerenZur:2010zw} of hydrodynamical anomaly.

One can expect that the polarization is proportional to the
anomalously induced axial current \cite{Son:2009tf}
\begin{equation}
  j^\mu_A \sim \mu^2 ~\left(1 \,-\, \frac{2~\mu ~ n}{3~(\epsilon+P)} \right)~
  \epsilon^{\mu\nu\lambda\rho}~ V_\nu ~\partial_{\lambda} V_\rho
\end{equation}
where $n$ and $\epsilon$ are the corresponding charge and energy
densities and $P$ is the pressure. Therefore, the $\mu$-dependence
of the polarization has to be more strong than that of CVE leading
to the effect rapidly increasing with decreasing energy.

This option may be explored in the framework of the program of
polarization  studies at NICA \cite{nicaWP} performed in the both collision points
as well as at the low-energy scan program at RHIC.

To collect the polarization data from different events one need to
supplement the production plane with a sort of orientation. For
this purpose one might use the left-right asymmetry of {\it
forward} neutrons as it was done at RHIC
\cite{Selyuzhenkov:2006fc,:2009txa} or another observable,
interesting by itself. The last comment regards handedness
\cite{Efremov:1992pe}, namely, the P-odd multiparticle momenta
correlation. Its exploration in heavy ion collisions provides a
way of orienting the event plane and collecting data for $\Lambda$
polarization and other P-odd observables.

Finally, let us mention the possibility \cite{nicaWP} to study
P-even angular distributions of dileptons
\cite{Bratkovskaya:1995kh} which might be used as probes of
quadratic effects of CME \cite{Buividovich:2010tn} and, quite
probably, CVE.

\section*{Acknowledgements}

We are indebted to D.B. Blaschke, P.V. Buividovich, A.V.~Efremov, P.~Fre,
V.D.~Kekelidze, D.E.~Kharzeev, R.~Lednicky, M.I.~Polikarpov,
V.D.~Toneev, V.I.~Zakharov, S.A.~Voloshin, K.R.~Mikhailov and Nu Xu for useful
discussions and comments. This work was supported in part by the
Russian Foundation for Basic Research (Grants No. 08-02-01003,
09-02-00732, 09-02-01149, 09-01-12179).

\end{document}